# Configurational Entropy and Its Scaling Behavior in Lattice Systems with Number of States Defined by Coordination Numbers.

Youshen Wu, Xin Guan, Shengli Zhang, Lei Zhang[*]

MOE Key Laboratory for Nonequilibrium Synthesis and Modulation of Condensed Matter, School of Physics, Xi'an Jiaotong University, Xi'an 710049, China

**ABSTRACT**. We introduce an exactly solvable lattice model that reveals a universal finite-size scaling law for configurational entropy driven purely by geometry. Using exact enumeration via Burnside's lemma, we compute the entropy for diverse 1D, 2D, and 3D lattices, finding that the deviation from the thermodynamic limit $s_\infty = \ln(z)$ scales as $\Delta s_N \sim N^{-1/d}$, with lattice-dependent higher-order corrections. This scaling, observed across structures from chains to FCC and diamond lattices, offers a minimal framework to quantify geometric influences on entropy. The model captures the order of magnitude of experimental residual entropies (e.g., $S_{\text{molar}} = R\ln 12 \approx 20.7 \,\text{J/mol} \cdot \text{K}$ for FCC plastic crystals) and provides a reference for understanding entropy-driven order in colloids, clusters, and solids.

## I. INTRODUCTION.

Entropy is a cornerstone of statistical physics, linking microscopic configurations to macroscopic thermodynamics [1]. Traditional lattice models often assume fixed degrees of freedom per site [2], yet real ordered systems exhibit rich entropy behaviors, such as residual entropy defying the third law and entropy-driven ordering. In ice, for instance, hydrogen bond arrangements yield a residual entropy of approximately 3.4 J/mol·K, as calculated by Pauling [3], with extensions to two-dimensional models highlighting boundary effects [4]. Plastic crystals maintain positional order but allow orientational disorder, resulting in high residual entropy and plasticity, while colloidal systems form ordered structures through entropy maximization, as seen in hard-particle assemblies [5]. These phenomena underscore the pivotal role of local geometry, particularly nearest-neighbor coordination [6,7]: surface atoms show enhanced catalysis due to undersaturation [8], nanoparticles are dominated by surface-to-volume ratios [9], local heterogeneities shape macroscopic properties in plastic crystals and disordered systems [10,11], and coordination governs self-assembly stability in colloids [12].

Can local geometry's impact on configurational entropy be isolated without energy interactions? We address this by introducing a lattice model where each site's state count equals its coordination number, providing a parameter-free geometric framework. In real systems, coordination modulates molecular or colloidal dynamics—high values (e.g., 12 in FCC) enable greater orientational freedom, yielding transition entropies around 20 J/mol·K [13,14]. Our model quantifies this effect purely from geometry. We develop a Burnside's lemma-based numerical solver to compute entropy across 1D-3D lattices, analyze finite-size scaling under open boundary conditions (OBC) revealing power-law decay from surface-volume effects, and use periodic boundary conditions (PBC) to confirm geometric origins. This framework elucidates entropy-driven processes in plastic crystals and colloids, offering insights into residual entropy, assembly stability, and geometric influences in ordered systems.

## II. METHODS

The lattice system studied in this paper consists of $N$ lattice sites in a $d$-dimensional lattice. For any site $i$, its number of states is defined as the nearest-neighbor coordination number $n_i$, meaning each site has a number of possible states equal to its local geometric connectivity, as illustrated in Figure 1, which depicts the lattice system. This assumption establishes the most direct, parameter-free relationship between local geometry and a site's state space, serving as an ideal zero-order model to isolate the purely geometric contributions to entropy.

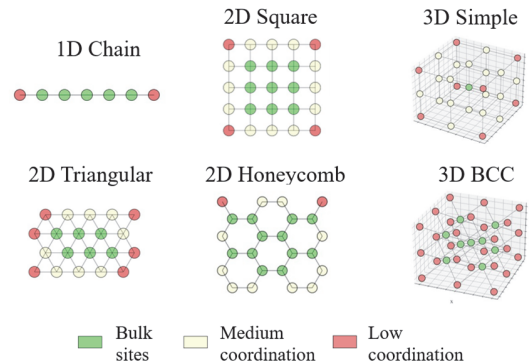

FIG 1. Lattice system with number of states defined by coordination numbers.

The lattice structures covered in this paper include the following 8 typical systems: one-dimensional: linear chain (bulk coordination number z=2, 1D-C); two-dimensional: square lattice (z=4, 2D-SQ), triangular lattice (z=6, 2D-TRI), honeycomb lattice (z=3, 2D-HC); three-dimensional: simple cubic (z=6, 3D-SC), diamond structure (z=4, Diamond), body-centered cubic (z=8, BCC), face-centered cubic (z=12, FCC).

*Contact author: zhangleio@mail.xjtu.edu.cn

In the model, all configurations are assumed to be completely degenerate in energy, with no interaction energies or preference ordering. Without considering the overall symmetry of the system, the total number of labeled configurations is $W_{labeled} = \prod_{i=1}^{N} n_i$, The number of physically distinct configurations is denoted as $\Omega$. The configurational entropy $S$ of the system is defined as $S = \ln\Omega$. For ease of comparison across lattice structures and dimensions, this paper primarily analyzes the dimensionless per-site configurational entropy:

$$s_N = \ln\Omega/N. \quad (1)$$

In the thermodynamic limit $N \to \infty$, the boundary effects of the system vanish, and the main contribution to entropy comes from interior sites. Considering symmetry factors, the actual number of configurations satisfies the inequality:

$$\frac{W_{labeled}}{|G|} \leq \Omega \leq W_{labeled}, \quad (2)$$

where $|G|$ is the order of the symmetry group corresponding to the lattice. Taking the logarithm of the above inequality and dividing by $N$ yields:

$$\frac{\ln W_{labeled} - \ln|G|}{N} \leq \frac{\ln\Omega}{N} \leq \frac{\ln W_{labeled}}{N}. \quad (3)$$

Since $|G|$ is a finite constant, in the limit $N \to \infty$, the upper and lower bounds converge, indicating that the limit of the per-site entropy exists and is:

$$s_\infty = \lim_{N \to \infty} \frac{1}{N} \sum_{i=1}^{N} \ln(n_i). \quad (4)$$

Further, decomposing the total configurational entropy into contributions from bulk and boundary sites:

$$\sum_{i=1}^{N} \ln(n_i) = \sum_{i \in \text{bulk}} \ln(n_i) + \sum_{i \in \text{boundary}} \ln(n_i), \quad (5)$$

where $z$ is the bulk coordination number, and $z'$ is the average coordination number of boundary sites. Since in a $d$-dimensional system, the number of boundary sites $N_{\text{boundary}} \sim N^{(d-1)/d}$, while $N_{\text{bulk}} \sim N$, the boundary term vanishes in the limit, ultimately yielding:

$$s_\infty = \ln(z_{bulk}). \quad (6)$$

For finite-size systems, the configurational entropy deviation $\Delta s = s_\infty - s_N$ primarily arises from boundary effects. Since the proportion of boundary sites is $\sim N^{-1/d}$, it can be inferred that the entropy deviation obeys the following power-law scaling relation:

$$\Delta s_N \sim N^{-1/d}, \quad (7)$$

This relation is consistent with the classic "surface-volume effect" in finite-size scaling theory [15,16].

To efficiently and accurately compute the system's configurational entropy $S = \ln\Omega$, this paper constructs a logarithmic-space numerical solver based on Burnside's lemma. Burnside's lemma is an important component of Pólya enumeration theory, with wide applications in graph theory, chemical counting problems, and statistical physics [17]. Its basic formula is:

$$\Omega = \frac{1}{|G|} \sum_{g \in G} \text{Fix}(g), \quad (8)$$

where $G$ is the symmetry group of the system, and $\text{Fix}(g)$ is the number of labeled configurations fixed under operation $g$. To avoid exponential overflow issues, this formula is computed in logarithmic space:

$$\ln\Omega = -\ln|G| + \ln\left(\sum_{g \in G} \exp(\ln|\text{Fix}(g)|)\right), \quad (9)$$

employing the log-sum-exp technique for numerical stability.

To validate the correctness of the algorithm, exhaustive comparisons were performed on several small-sized systems, with results in complete agreement, as shown in Table 1, demonstrating the algorithm's high accuracy and generality.

TABLE I. Exhaustive Validation of the Algorithm.

| Lattice | Size (N) | $\Omega$ | $\Omega$ (Exhaustive) |
|---|---|---|---|
| 1D-C | 4 | 3 | 3 |
| 2D-SQ | 4 | 6 | 6 |
| 2D-TRI | 4 | 21 | 21 |
| 2D-HC | 8 | 78 | 78 |
| 3D-SC | 8 | 267 | 267 |
| FCC | 4 | 15 | 15 |
| BCC | 9 | 8 | 8 |
| Diamond | 17 | 140 | 140 |

## III. RESULTS

Using the solver constructed based on Burnside's lemma, systematic calculations were performed for the eight lattice structures under open boundary conditions (OBC). Figure 2 shows the trend of per-site entropy $s_N$ with increasing system size $N$. The $s_N$ for all systems monotonically approaches the theoretical limit $s_\infty = \ln(z)$ from below.

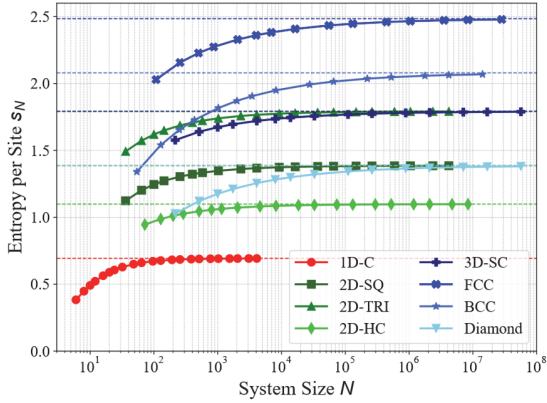

FIG 2. Convergence behavior of per-site entropy $s_N$ with system size $N$ for various lattices in different dimensions (semi-logarithmic coordinates). Colored dashed lines represent the theoretical limits $\ln(z)$ for each lattice.

To further quantitatively analyze the convergence behavior, the relationship between entropy deviation $\Delta s = s_\infty - s_N$ and system size $N$ is examined. Figure 3 shows the correspondence between $\Delta s$ and $N$ in double-logarithmic coordinates, with all lattice data aligning along straight lines, verifying the power-law scaling behavior:

$$\Delta s_N \sim N^{-\alpha}, \quad \alpha \approx \frac{1}{d}. \tag{10}$$

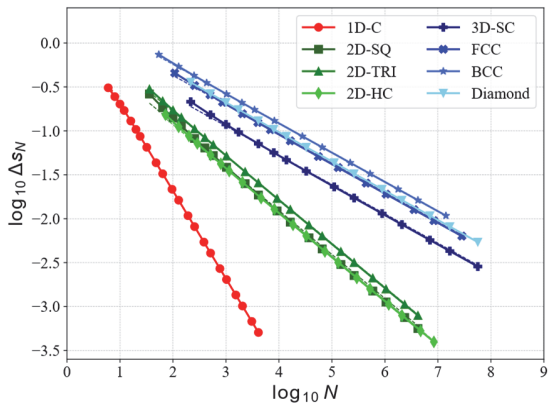

FIG 3. Double-logarithmic plot of entropy deviation $\Delta s_N$ versus system size $N$. All lattices exhibit strong linear relationships, with slopes highly consistent with the theoretical value $-1/d$.

Linear fitting was performed on the data in the figure, and the scaling exponents $\alpha$ extracted for each lattice system are summarized in Table 2, showing high consistency with the theoretical value $1/d$, further confirming the universality of this scaling law.

TABLE II. Various Fitted Scaling Exponents $\alpha$

| Lattice | Dimension(d) | 1/d | Fitted α |
|---|---|---|---|
| 1D-C | 1 | 1.000 | 1.000 |
| 2D-SQ | 2 | 0.500 | 0.521 |
| 2D-TRI | 2 | 0.500 | 0.507 |
| 2D-HC | 2 | 0.500 | 0.508 |
| 3D-SC | 3 | 0.333 | 0.343 |
| FCC | 3 | 0.333 | 0.340 |
| BCC | 3 | 0.333 | 0.337 |
| Diamond | 3 | 0.333 | 0.335 |

To verify the physical origin of entropy deviation, periodic boundary conditions (PBC) were introduced in typical two-dimensional (square lattice) and three-dimensional (SC) lattices (while taking into account the symmetry group changes caused by PBC) for comparative analysis. Figure 4 shows that, at the same system size, the per-site entropy under PBC is significantly closer to its theoretical limit, with a markedly improved convergence speed. This result clearly indicates that the dominant correction to configurational entropy in finite sizes originates from boundary effects, further corroborating the "surface-volume" driven scaling mechanism.

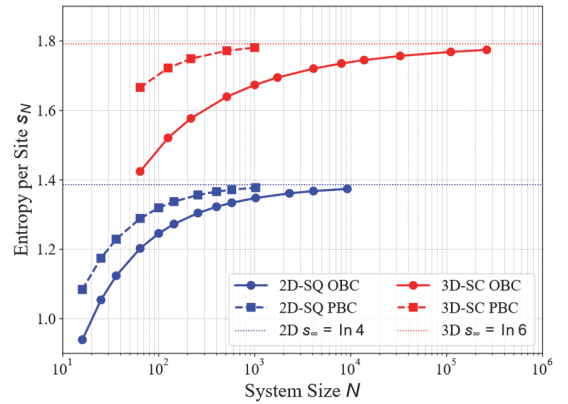

FIG 4. Comparison of per-site entropy convergence under OBC and PBC for 2D square and 3D simple cubic lattices.

## IV. DISCUSSION

The purely geometric model constructed in this study not only reveals the dominant influence of local coordination on configurational entropy but also verifies a finite-size scaling law. Its deeper physical implications can be explored from three perspectives: the universality of the scaling law, the model's physical applications, and future extensions.

The most central finding of this research is that the configurational entropy deviation follows the power-law scaling relation $\Delta s_N \sim N^{-1/d}$. This law holds across various lattice types from one to three dimensions, powerfully demonstrating that in systems without energy interactions, finite-size effects are primarily determined by the fundamental geometric factor of the surface-to-volume ratio.

More interestingly, the fitted scaling exponent $\alpha$ is systematically slightly larger than the theoretical value $1/d$. This minor deviation is not an error but a manifestation of the influence of higher-order geometric features, such as edges and corners. Since the complete expansion of the entropy is:

$$\Delta s_N = C_1 N^{-1/d} + C_2 N^{-2/d} + \cdots, \quad (10)$$

the presence of higher-order terms increases the effective exponent [18]. The differences in the degree of deviation among various lattices (e.g., square vs. triangular, simple cubic vs. BCC) indicate that the precise value of the scaling exponent can serve as a sensitive "geometric fingerprint" that quantitatively reflects the subtle geometric differences of various lattice structures.

By defining the number of states at each site by its coordination number $n_i$, this model provides an ideal geometric reference framework for quantifying the phenomenon of "entropy-driven order." In systems with weak energy interactions, such as plastic crystals and colloids, particles achieve an increase in overall entropy and form ordered structures by maximizing free volume or orientational freedom through specific geometric arrangements [19, 20]. A higher coordination number $n_i$ implies weaker local constraints and more available microstates to explore, thus leading to higher configurational entropy. The model offers a geometric explanation for the high transition entropy of plastic crystals. For example, for an FCC lattice with a high coordination number ($z = 12$), the model predicts a molar configurational entropy of $S_{molar} = R\ln(12) \approx 20.66$ J/mol·K. This value is on the same order of magnitude as the experimental transition entropies of many FCC plastic crystals (18–36 J/mol·K) [13, 14, 21], indicating that even while ignoring energy and vibrational contributions, the model still captures the core contribution to entropy from the orientational freedom afforded by high coordination.

Under the free energy approximation $F \approx -TS_{config}$, system stability is determined by maximizing the configurational entropy $S_{config} \approx k_B \sum \ln n_i$. The scaling law from this study quantitatively describes the "entropic surface penalty" caused by low-coordination surface particles: $\Delta F_{surface} \approx k_B T \cdot N \cdot \Delta s_N \sim N^{(d-1)/d}$. This directly explains the stability of "magic-number" clusters in colloidal self-assembly—these clusters achieve the lowest overall free energy by forming complete geometric shells, which maximizes surface coordination and thus minimizes the entropic penalty [22]. This idea is also consistent with the concept of engineering inter-particle forces via "shape entropy" [23]. Furthermore, using this entropy model as an optimization objective also provides an effective pathway for the inverse design of colloidal crystals [24].

The model's greatest advantage is its ultimate simplicity, which makes it easy to analyze and allows it to reveal universal laws. However, this also means it cannot describe energy-driven phase transitions. Future research could be extended in the following directions: Introduce energy weights into the number of states to construct a complete Boltzmann statistical model, or generalize the state function to include next-nearest neighbors, $f(n_{i1}, n_{i2})$, to enhance the model's explanatory power. The framework can be applied to study disordered systems, such as entropy-driven disordered clusters caused by symmetry breaking [25], or the spatial correlation between configurational entropy and local structures in glassy materials [26]. An interesting direction would be to apply this model's local coordination entropy calculation ($S_{config} \sim \sum \ln n_i$) to the various random tiling configurations of quasicrystals, such as those recently studied by Fayen et al. [27], to test whether a structure's local entropy is correlated with its global tiling entropy.

## V. CONCLUSIONS

We introduce a geometric lattice model where site states are defined by nearest-neighbor coordination number, isolating geometry's role in configurational entropy without energy interactions. Employing Burnside's lemma, we exactly compute entropy for diverse lattices: 1D chains; 2D square, triangular, and honeycomb; and 3D simple cubic, face-centered cubic, body-centered cubic, and diamond structures. Results show per-site entropy $s_N$ monotonically converges from below to $\ln(z)$ (with bulk coordination $z$) and follows finite-size scaling $\Delta s_N \sim N^{-1/d}$, with lattice-dependent variations reflecting higher-order geometric corrections from edges and corners. Comparisons of open and periodic boundaries confirm surface-volume origins of these effects in noncritical systems.

This parameter-free model provides a solvable paradigm for quantifying geometric entropy contributions, yielding insights into entropy-driven order in plastic crystals and colloidal assemblies, residual entropy in ordered solids, and self-assembly stability. It establishes a foundation for advancing nanoscale material design, disordered systems, and thermodynamic behaviors influenced by geometry, with extensions to energy or disorder poised to expand its scope.

### ACKNOWLEDGMENTS
We acknowledge support from the National Natural Science Foundation of China (NSFC Nos. 52273092 and T2425029). We thank Ma Zhi and Gu Xiang for valuable discussions.

### Data and Code Availability
The Python code used for calculations and plotting is publicly available to ensure transparency and reproducibility.